\documentclass
[preprint,showkeys,byrevtex]{revtex4} 
\usepackage{graphicx}
\usepackage{bm}
\begin{document}   
\preprint{CYCU-HEP-09-20}
\title{Chiral magnetic effect (CME) at low temperature\\
 from instanton vacuum}      
%-------------------------------------------------
\author{Seung-il Nam}
\email[E-mail: ]{sinam@cycu.edu.tw}
\affiliation{Department of Physics, Chung-Yuan Christian University (CYCU), 
Chung-Li 32023, Taiwan} 
%-------------------------------------------------
\date{\today}
\begin{abstract}  
In this talk, we report our present work on the chiral magnetic effect (CME) under a strong magnetic field $\bm{B}=B_{0}\hat{x}_{3}$ at low temperature. To this end, we use the instanton vacuum with the finite instanton-number fluctuation $\Delta$, which relates to the nontrivial topological charge $Q_{\mathrm{t}}$. We compute the vacuum expectation values of the local chiral density $\langle \rho_{\chi}\rangle$, chiral charge density $\langle n_{\chi}\rangle $ and induced electromagnetic current $\langle j_{\mu}\rangle$. We observed that the longitudinal EM current is much larger than the transverse one, $|j_{\perp}/j_{\parallel}|\sim Q_{\mathrm{t}}$, and the $\langle n_{\chi}\rangle $ equals to the $|\langle j_{3,4}\rangle|$. It also turns out that the CME becomes insensitive to the magnetic field as $T$ increases, since the instanton effect decreases. 
\end{abstract} 
\maketitle
%--------------------------------------------------
\section{Introduction}
%--------------------------------------------------
Recently, in Refs.~\cite{Kharzeev:2004ey,Voloshin:2004vk}, it was suggested that the nontrivial topological charge of QCD vacuum, $Q_{\mathrm{t}}$ can be probed by measuring asymmetric electric currents, due to an event-by-event $P$- and $CP$-violation, in the non-central heavy-ion collision experiments such as RHIC, FAIR, and LHC. Their main idea is that unequal numbers of the chirally left- and right-handed quarks, $N_{\mathrm{L}}-N_{\mathrm{R}}\ne0$ proportional to the $Q_{\mathrm{t}}$ in terms of the axial Ward-Takahashi identity~\cite{Kharzeev:2007jp}, can produce the asymmetric electric current under a strong magnetic field $B$. Interestingly enough, a possible experimental evidence was already reported by STAR collaboration at RHIC~\cite{Voloshin:2008jx}. Moreover, in Ref.~\cite{Buividovich:2009wi}, using the lattice QCD simulation, the electromagnetic (EM) current fluctuation as well as local chirality were explored at finite $T$, resulting in that the fluctuations become insensitive to the external magnetic field as $T$ increases. In Ref.~\cite{Fukushima:2008xe}, it was also shown that the electric current, induced by the external magnetic field with the nontrivial $Q_{\mathrm{t}}$, reads 
%EQUATION>>>
\begin{equation}
\label{eq:EMC}
{\bm j}=-\frac{\mu_{\chi}}{2\pi^{2}}{\bm B},
\end{equation}
%EQUAITON<<<
where the $\mu_{\chi}$ represents the chiral chemical potential and contains the information on the $Q_{\mathrm{t}}$ inside of it. In fact, this is a consequence of the EM axial anomaly~\cite{DHoker:1985yb}.

In the present report, based on our previous work~\cite{Nam:2009jb}, we investigate the CME at low $T$, employing the instanton vacuum configuration with the nontrivial $Q_{\mathrm{t}}$, which relates to the finite instanton-number fluctuation, i.e.  the number difference between the instantons and anti-instantons in the grand canonical ensemble, $\Delta\equiv N_{I}-N_{\bar{I}}\ne0$ resulting in the $CP$-violation. To this end, we first write the effective action derived from the instanton vacuum configuration as a functional of $\Delta$~\cite{Diakonov:1995qy}. Using generic functional external-source and linear Schwinger methods, we compute the relevant physical quantities, the vacuum expectation values of the local chiral density $\langle \rho_{\chi}\rangle$, chiral charge density $\langle n_{\chi}\rangle$, EM current $\langle j_{\mu} \rangle$, induced by the external magnetic field. In order to consider the $T$-dependence of the relevant quantities, we also employ the Harrington-Shepard caloron to obtain the $T$-dependent constituent-quark mass. Fianlly, the standard fermionic Matsubara formula is taken into account for the anti-perioidc sum over the Euclidean time, i.e.  $T$. We consider all the relevant physical quantities up to $\mathcal{O}(\Delta^{2})\sim\mathcal{O}(Q^{2}_{\mathrm{t}})$.

%--------------------------------------------------
\section{Effective action from the instanton vacuum}
%--------------------------------------------------
In this Section, we briefly introduce a $P$- and $CP$-violating effective action $\mathcal{S}_{\mathrm{eff}}$, derived by Diakonov {\it et al.} from the instanton vacuum configuration in the large $N_{c}$ limit at zero temperature ($T=0$)~\cite{Diakonov:1995qy}. Employing a dilute grand canonical ensemble of the (anti)instantons with a finite instanton-number fluctuation, $\Delta\equiv N_{+}-N_{-}\ne0$, which corresponds to a $CP$-violating vacuum, but a fixed total number of the pseudo-particles $N_{+}+N_{-}=N$, the $\mathcal{S}_{\mathrm{eff}}$ can be written in momentum space with Euclidean metric as follows:
%EQUATION>>>
\begin{eqnarray}
\label{eq:EA}
\mathcal{S}_{\mathrm{eff}}
&=&\mathcal{C}+\frac{N_{+}}{V}\ln\lambda_{+}
+\frac{N_{-}}{V}\ln\lambda_{-}
-\frac{mN_{c}}{4\pi^{2}\bar{\rho}^{2}}(\lambda_{+}+\lambda_{-})
\cr
&-&N_{c}\int\frac{d^{4}k}{(2\pi)^{4}}\mathrm{Tr}_{\gamma}\ln
\left[\frac{\rlap{/}{k}-\frac{i}{2}
[\lambda_{+}(1+\gamma_{5})+\lambda_{-}(1-\gamma_{5})]F^{2}(k)}
{\rlap{/}{k}-im}\right],
\end{eqnarray}
%EQUAITON<<<

From the effective action, we can obtain the following two self-consistent (saddle-point) equations with respect to the $\lambda_{\pm}$:
%EQUATION>>>
\begin{eqnarray}
\label{eq:SDP}
\lambda_{\pm}
\frac{N_{\pm}}{V}=\frac{(1\pm\delta)M_{0}mN_{c}}
{4\pi^{2}\bar{\rho}^{2}}
+N_{c}\int\frac{d^{4}k}{(2\pi)^{4}}\mathrm{Tr}_{\gamma}
\left[\frac{\frac{1}{2}(1\pm\gamma_{5})
(1+\delta\gamma_{5})^{2}M^2}
{k^{2}+(1+\delta\gamma_{5})^{2}M^2}\right],
\end{eqnarray}
%EQUAITON<<<
where the $\lambda_{\pm}$ is approximated as $M_{0}(1\pm\delta)$ in the last line of Eq.~(\ref{eq:SDP}) with account of the fact $\Delta\ll N$ in the thermodynamic limit~\cite{Diakonov:1995qy}. The momentum-dependent constituent-quark mass is defined as $M(k)=M_{0}F^{2}(k)$~\cite{Diakonov:2002fq}. By adding and subtracting the instanton $(+)$ and anti-instanton ($-$) contributions in Eq.~(\ref{eq:SDP}), we can obtain an expression for $\delta$ as a function of relevant parameters:
%EQUATION>>>
\begin{equation}
\label{eq:DEL}
\delta=\left(\frac{2\pi^{2}\bar{\rho}^{2}}{mM_{0}N_{c}}\right)
\frac{\Delta}{V}.
\end{equation}
%EQUAITON<<<
Taking into account all the ingredients discussed so far, finally, we can write the relevant effective action with $\Delta\ne0$ for further investigations:
%EQUATION>>>,
\begin{eqnarray}
\label{eq:EA2}
\mathcal{S}_{\mathrm{eff}}
&=&-\int\frac{d^4k}{(2\pi)^4}\mathrm{Tr}_{c,f,\gamma}\ln
\left[\frac{\rlap{/}{k}-i(1+\delta\gamma_{5})M(k)}
{\rlap{/}{k}-im}\right],
\end{eqnarray}
%EQUAITON<<<
where the $\mathrm{Tr}_{c,f,\gamma}$ denotes the trace over color, flavor  and Lorentz indices. 

According to the axial Ward-Takahashi identity~\cite{Kharzeev:2007jp}, the $Q_{\mathrm{t}}$ is proportional to the number difference between the chirally left- and right-handed quarks, $Q_{\mathrm{t}}\propto N_{R}-N_{L}$. Hence, the nontrivial $Q_{\mathrm{t}}$ indicates the chirality flip. Note that, similarly, if a chirally left-handed quark is scattered from an instanton to an anti-instanton, the quark helicity is flipped to the right-handed one, and vice versa. This means that the nonzero $\Delta$ results in $N_{R}-N_{L}\ne0$. In this way, the $Q_{\mathrm{t}}$ can be considered to be proportional to the $\Delta$: $Q_{\mathrm{t}}\sim\Delta$~\cite{Schafer:1996wv,Diakonov:2002fq}. The $\delta$ in Eq.~(\ref{eq:DEL}) can be rewritten in terms of a real and small parameter $\epsilon$, which satisfies the condition $|\epsilon|\leq1$, for convenience as follows:
%EQUATION>>>
\begin{equation}
\label{eq:44}
\delta=\left(\frac{2\pi^{2}\bar{\rho}^{2}}{mM_{0}N_{c}}\right)
\frac{\epsilon N}{V},
\end{equation}
%EQUAITON<<<

%--------------------------------------------------
\section{The Chiral magnetic effect in the magnetic field}
%-------------------------------------------------
In this Section, we compute relevant physical quantities for the CME in the presence of the external magnetic field, using the effective action given in the last Section. They are the vacuum expectation values (v.e.v.) of the local chiral density $\rho_{\chi}$, chiral charge density $n_{\chi}$, and electromagnetic (EM) current $j_{\mu}$, which are defined as follows:
%EQUATION>>>
\begin{equation}
\label{eq:DEF}
\rho_{\chi}(x)=iq^{\dagger}(x)\gamma_{5}q(x),\,\,\,\,
n_{\chi}(x)=iq^{\dagger}(x)\gamma_{4}\gamma_{5}q(x),\,\,\,\,
j_{\mu}(x)=iq^{\dagger}(x)\gamma_{\mu}q(x).
\end{equation}
%EQUAITON<<<
The $\rho_{\chi}$ and $n_{\chi}$ represent the strength of the parity breaking in a system, whereas the $j_{\mu}$ here the EM current induced by the external magnetic field. It is very convenient to compute these quantities employing the effective action with corresponding pseudoscalar ($\mathcal{P}$), axial vector ($\mathcal{A}$), and  vector ($\mathcal{V}$) external sources. Hence, the effective action can be rewritten as: 
%EQUATION>>>
\begin{equation}
\label{eq:EAV}
\mathcal{S}_{\mathrm{eff}}
=-\int\frac{d^4k}{(2\pi)^4}\mathrm{Tr}\ln
\left[\frac{\rlap{/}{K}-i(1+\delta\gamma_{5})M(K)
+i\delta\gamma_{5}\mathcal{P}
+\delta\gamma_{4}\gamma_{5}\mathcal{A}_{4}
+\gamma_{\mu}\mathcal{V}_{\mu}}
{\rlap{/}{K}-im+i\delta\gamma_{5}\mathcal{P}
+\delta\gamma_{4}\gamma_{5}\mathcal{A}_{4}
+\gamma_{\mu}\mathcal{V}_{\mu}}\right],
\end{equation}
%EQUAITON<<<
where $\mathrm{Tr}\equiv\mathrm{Tr}_{c,f,\gamma}$ with $N_{c}=3$ and $N_{f}=2$. The $K_{\mu}$ indicates the covariant quark momentum, gauged by the photon field as $k_{\mu}+A_{\mu}$. Note that we have written the external sources $\mathcal{P}$ and $\mathcal{A}_{4}$, which correspond to the $\rho_{\chi}$ and $n_{\chi}$, to be proportional to the $\delta$, since they signal the nontrivial topological charge $Q_{\mathrm{t}}$ as the parity-breaking quark mass $\delta\gamma_{5}M$ does.  
 
First, we calculate the v.e.v. of the chiral density, using the standard functional method as follows: 
%EQUATION>>>
\begin{equation}
\label{eq:CD1}
\langle\rho_{\chi}\rangle
=-i\delta N_{c}N_{f}\int\frac{d^{4}k}{(2\pi)^{4}}\mathrm{Tr}_{\gamma}
\left\{\left[
\frac{1}{\rlap{/}{K}-i(1+\delta\gamma_{5})M(K)}
-\frac{1}{\rlap{/}{K}-im}\right]\gamma_{5} \right\},
\end{equation}
%EQUAITON<<<
where we have performed the trace of over the color and flavor indices. Since we are interested in the response of the nonperturbative vacuum to  the external magnetic field for the nonzero topological charge $Q_{\mathrm{t}}$, we only collect the terms proportional to the $\delta$ and the photon field strength tensor $F_{\mu\nu}$. In order to perform the trace over the Lorentz index in the r.h.s. of Eq.~(\ref{eq:CD1}) under the external EM field, we employ the linear Schwinger method~\cite{Schwinger:1951nm,Nieves:2006xp}. According to that, the quark propagator in the presence of the instanton background can be written up to $\mathcal{O}(F_{\mu\nu})$, which is equivalent up to $\mathcal{O}(B_{0})$, as follows:
%EQUATION>>>
\begin{eqnarray}
\label{eq:DEEX}
S(k,A)&=&\frac{1}{\rlap{/}{K}-i(1+\delta\gamma_{5})M(K)}
\approx
\frac{\rlap{/}{k}+\rlap{/}{A}
+i(1+\delta\gamma_{5})\left[M+\frac{1}{2}\bar{M}
(\sigma\cdot F)\right]}
{k^{2}+(1+\delta^{2})M^{2}}
\cr
&\times&
\left[1-\frac{\tilde{M}(\sigma\cdot F)+i\left(1-\frac{\delta}{4}\gamma_{5} \right)\hat{M}(k)\gamma_{\mu}K_{\nu}F_{\mu\nu}-2i\delta M\gamma_{5}\rlap{/}{K}}{k^{2}+(1+\delta^{2})M^{2}}\right].
\end{eqnarray}
%EQUAITON<<<
Here, $\sigma\cdot F=\sigma_{\mu\nu}F_{\mu\nu}$. We define relevant functions related to the momentum-dependent quark mass:
%EQUATION>>>
\begin{equation}
\label{eq:MF}
M=M_{0}\left(\frac{2}{2+k^{2}\bar{\rho}^{2}} \right)^{2},\,\,\,\,
\bar{M}=-\frac{8M_{0}\bar{\rho}^{2}}{(2+k^{2}\bar{\rho}^{2})^{3}},\,\,\,\,
\tilde{M}=\frac{1}{2}+M\bar{M},\,\,\,\,\hat{M}=4i\bar{M}.
\end{equation}
%EQUAITON<<<
After a straightforward trace manipulation, one arrives at a simple expression for the chiral density:
%EQUATION>>>
\begin{equation}
\label{eq:CD2}
\langle\rho_{\chi} \rangle_{F,\delta}
=2\delta^{2}\mathcal{F}_{a}B^{2}_{0},
\end{equation}
%EQUAITON<<<
where the subscripts $F$ and $\delta$ in the l.h.s. of Eq.~(\ref{eq:CD2}) indicate the fact that we picked up only the terms proportional to the $F_{\mu\nu}$ and $\delta$ as explained above. In deriving Eq.~(\ref{eq:CD2}), to make the problem easy, we assumed a static external EM field, which is assigned for example as in Ref.~\cite{Buividovich:2009wi}: 
%EQUATION>>>
\begin{equation}
\label{eq:A}
A_{\mu}=A^{\mathrm{cl}}_{\mu}+A^{\mathrm{fluc}}_{\mu}
=\left(-\frac{B_{0}}{2}x_{2},\frac{B_{0}}{2}x_{1},0,0 \right)+(a_{1},a_{2},a_{3},a_{4}),
\end{equation}
%EQUAITON<<<
where the external EM field consists of the classical ($A^{\mathrm{cl}}_{\mu}$) in the symmetry gauge and fluctuation ($A^{\mathrm{fluc}}_{\mu}$) parts. We take the $a_{1\sim4}$ as small and constant EM potentials. Among the EM field strength tensors, $F_{12}=B_{0}$ and its dual remain finite, while others disappear. The relevant integral $\mathcal{F}_{a}$ in Eq.~(\ref{eq:CD2}) reads:
%EQUATION>>>
\begin{equation}
\label{eq:Fa}
\mathcal{F}_{a}=-4N_{c}N_{f}\int\frac{d^{4}k}{(2\pi)^{4}}
\frac{\bar{M}\left(M\bar{M}+\frac{1}{2} \right)}
{[k^{2}+(1+\delta^{2})M^{2}]^{2}}, 
\end{equation}
%EQUAITON<<<
As understood from Eq.~(\ref{eq:CD2}), the parity breaking of the vacuum becomes enhanced quadratically with respect to the $\delta$ and $B_{0}$.

Similarly to the $\rho_{\chi}$, we can compute the v.e.v. of the chiral charge density as follows:
%EQUATION>>>
\begin{equation}
\label{eq:CCD1}
\langle{n_{\chi}}\rangle=
-\delta N_{c}N_{f}\int\frac{d^{4}k}{(2\pi)^{4}}\mathrm{Tr}_{\gamma}
\left\{\left[
\frac{1}{\rlap{/}{K}-i(1+\delta\gamma_{5})M(K)}
-\frac{1}{\rlap{/}{K}-im}
\right]\gamma_{4}\gamma_{5}\right\}.
\end{equation}
%EQUAITON<<<
By solving Eq.~(\ref{eq:CCD1}) and picking up relevant terms, one is lead to a compact expression for it as follows:
%EQUATION>>>
\begin{equation}
\label{eq:CCD2}
\langle{n_{\chi}}\rangle_{F,\delta}
=\delta\mathcal{F}_{b}\epsilon_{4\nu\rho\sigma}(iA_{\sigma})F_{\nu\rho}
=2\delta\mathcal{F}_{b}(iA_{3})B_{0},
\end{equation}
%EQUAITON<<<
where we have used the fact that only the $\tilde{F}_{34}=B_{0}$ is nonzero in our symmetric-gauge EM field. From Eq.~(\ref{eq:CCD2}), we observe that the chiral charge density increases linearly with respect to the $\delta$ as well as the $B_{0}$. The integral $\mathcal{F}_{b}$ is written as: 
%EQUATION>>>
\begin{equation}
\label{eq:Fb}
\mathcal{F}_{b}=-4N_{c}N_{f}\int\frac{d^4k}{(2\pi)^4}
\frac{M\bar{M}}
{[k^{2}+(1+\delta^{2})M^{2}]^{2}}.
\end{equation}
%EQUAITON<<<
Finally, we attempt to compute the v.e.v. of the EM current, induced by the external magnetic field in the presence of the nontrivial topological charge:
%EQUATION>>>
\begin{equation}
\label{eq:EMC1}
\langle{j_{\mu}}\rangle=-N_{c}N_{f}\int\frac{d^{4}k}{(2\pi)^{4}}
\mathrm{Tr}_{\gamma}
\left\{\left[
\frac{1}{\rlap{/}{K}-i(1+\delta\gamma_{5})M(K)}
-\frac{1}{\rlap{/}{K}-im}
\right]\gamma_{\mu}\right\}.
\end{equation}
%EQUAITON<<<
Expectedly, we can obtain a very similar expression for the $\langle j_{\mu}\rangle_{F,\delta}$ to that for the v.e.v. of the chiral charge density:
%EQUATION>>>
\begin{eqnarray}
\label{eq:EMC2}
\langle{j_{\mu}}\rangle_{F,\delta}=
-[3\delta^{2}F_{\mu\nu}+2\delta\tilde{F}_{\mu\nu}](iA_{\nu})\mathcal{F}_{b},
\end{eqnarray}
%EQUAITON<<<
where the definition of the $\mathcal{F}_{b}$ is the same with that given in Eq.~(\ref{eq:Fb}). Considering Eq.~(\ref{eq:EMC2}) and assuming finite values for $A_{1\sim4}$, we can write the induced EM currents for $Q_{\mathrm{t}}\ne0$ separately for its transverse ($\perp$) and longitudinal ($\parallel$) components: 
%EQUATION>>>
\begin{eqnarray}
\label{eq:JOA}
\langle{j_{1,2}}\rangle_{F,\delta}
&\equiv&
j_{\perp}=\mp3\delta^{2}\mathcal{F}_{b}(iA_{2,1})B_{0},
\cr
\langle{j_{3,4}}\rangle_{F,\delta}
&\equiv&
j_{\parallel}=\mp2\delta\mathcal{F}_{b}(iA_{4,3})B_{0}.
\end{eqnarray}
%EQUAITON<<<
The $j_{\parallel}$ increases linearly as the $B_{0}$ grows, which is the indication of the CME as in Eq.~(\ref{eq:EMC}). Equating the transverse and longitudinal components, we have the ratio as follows:
%EQUATION>>>
\begin{equation}
\label{eq:JJJ}
\left|\frac{j_{\perp}}{j_{\parallel}}\right|=\frac{3}{2}\frac{|A_{2,1}|}{|A_{4,3}|}\delta.
\end{equation}
%EQUAITON<<<
From Eq.~(\ref{eq:JJJ}), if we assume that $|A_{1,2}|\sim|A_{3,4}|$ along the $\hat{x}_{3}$-direction, in other words, the strengths of the fluctuations $a_{1\sim4}$ in Eq.~(\ref{eq:A}) are almost the same to each other, it is obvious that the transverse components of the induced EM current are much smaller than those of the longitudinal ones, $|j_{\perp}/j_{\parallel}|\ll1$, due to the fact that $\delta\ll1$. This tendency is just consistent with that given in the recent lattice QCD simulation~\cite{Buividovich:2009wi}.

Considering the general expression for the induced EM current in Eq.~(\ref{eq:EMC}), the induced EM current along the $\hat{x}_{3}$-direction in Eq.~(\ref{eq:JOA}) can be rewritten as
%EQUATION>>>
\begin{equation}
\label{eq:JJJ1}
\langle j_{3}\rangle_{F,\delta}=
-\frac{1}{2\pi^{2}}[4\pi^{2}\delta\mathcal{F}_{b}(iA_{4})]B_{0},
\end{equation}
%EQUAITON<<<
and gives the following expression for the chiral chemical potential,
%EQUATION>>>
\begin{equation}
\label{eq:JJJ2}
\mu_{\chi}
=4\pi^{2}\mathcal{F}_{b}(i\delta A_{4}).
\end{equation}
%EQUAITON<<<
Comparing Eqs.~(\ref{eq:CCD2}) and (\ref{eq:JOA}), it can be easily shown that the v.e.v. of the chiral charge density is the same with the third and fourth components of the induced EM current in the leading contributions in $\delta$:
%EQUATION>>>
\begin{equation}
\label{eq:NA}
\langle n_{\chi} \rangle_{F,\delta}=\mp\langle j_{3,4} \rangle_{F,\delta}
\approx\frac{1}{2\pi^{2}}(i\delta A_{4,3})B_{0}
=\frac{1}{2\pi^{2}}\mu_{\chi}B_{0},
\end{equation}
%EQUAITON<<<
where we have used the result from Eq.~(\ref{eq:JJJ2}). Eq.~(\ref{eq:NA}) for $A_{3}$ is consistent with that given in Refs.~\cite{Sadooghi:2006sx,Fukushima:2008xe} for a homogeneous magnetic field.

%--------------------------------------------------
\section{Chiral magnetic effect at low temperature}
%--------------------------------------------------
To investigate the physical quantities in hand at low but finite temperature ($T\lesssim T^{\chi}_{c}$), we want to discuss briefly how to modify the instanton variables, $\bar{\rho}$ and $\bar{R}$ at finite $T$. We will follow our previous work~\cite{Nam:2009nn} and Refs.~\cite{Harrington:1976dj,Diakonov:1988my} to this end. We write the instanton distribution function at finite $T$ with the HS caloron as follows:
%EQUATION>>>
\begin{equation}
\label{eq:IND}
d(\rho,T)=\underbrace{C_{N_c}\,\Lambda^b_{\mathrm{RS}}\,
\hat{\beta}^{N_c}}_\mathcal{C}\,\rho^{b-5}
\exp\left[-(A_{N_c}T^2
+\bar{\beta}\gamma n\bar{\rho}^2)\rho^2 \right].
\end{equation}
%EQUATION<<<
Using the instanton distribution function in Eq.~(\ref{eq:IND}), we can compute the average value of the instanton size, $\bar{\rho}^2$ straightforwardly as follows~\cite{Schafer:1996wv}:
%EQUATION>>>
\begin{equation}
\label{eq:rho}
\bar{\rho}^2(T)
=\frac{\int d\rho\,\rho^2 d(\rho,T)}{\int d\rho\,d(\rho,T)}
=\frac{\left[A^2_{N_c}T^4
+4\nu\bar{\beta}\gamma n \right]^{\frac{1}{2}}
-A_{N_c}T^2}{2\bar{\beta}\gamma n},
\end{equation}
%EQUATION<<<
where $\nu=(b-4)/2$. Substituting Eq.~(\ref{eq:rho}) into Eq.~(\ref{eq:IND}), the distribution function can be evaluated further as:
%EQUATION>>>
\begin{equation}
\label{eq:dT}
d(\rho,T)=\mathcal{C}\,\rho^{b-5}
\exp\left[-\mathcal{M}(T)\rho^2 \right],\,\,\,\,
\mathcal{M}(T)=\frac{1}{2}A_{N_c}T^2+\left[\frac{1}{4}A^2_{N_c}T^4
+\nu\bar{\beta}\gamma n \right]^{\frac{1}{2}}.
\end{equation}
%EQUATION<<<

Finally, in order to estimate the $T$-dependence of the constituent-quark mass $M_{0}$, it is necessary to consider the normalized distribution function, defined as follows:
%EQUATION>>>
\begin{equation}
\label{eq:NID}
d_N(\rho,T)=\frac{d(\rho,T)}{\int d\rho\,d(\rho,T)}
=\frac{\rho^{b-5}\mathcal{M}^\nu(T)
\exp\left[-\mathcal{M}(T)\rho^2 \right]}{\Gamma(\nu)}.
\end{equation}
%EQUATION<<<
Now, we want to employ the large-$N_c$ limit to simplify the expression of $d_N(\rho,T)$. Since the parameter $b$ becomes infinity as $N_c\to\infty$, and the same for $\nu$. In this limit, as understood from Eq.~(\ref{eq:NID}), $d_N(\rho,T)$ can be approximated as a $\delta$-function~\cite{Diakonov:1995qy}: 
%EQUATION>>>
\begin{equation}
\label{eq:NID2}
\lim_{N_c\to\infty}d_N(\rho,T)=\delta[{\rho-\bar{\rho}\,(T)}].
\end{equation}
%EQUATION<<<
Considering the constituent-quark mass is represented by~\cite{Diakonov:1995qy} 
%EQUATION>>>
\begin{equation}
\label{eq:M0}
M_{0}\propto\sqrt{n}\int d\rho\,\rho^{2}\delta[\rho-\bar{\rho}(T)]
=\sqrt{n(T)}\,\bar{\rho}^{2}(T),
\end{equation}
%EQUAITON<<<
we can modify the $M_{0}$ as a function of $T$ as follows:
%EQUATION>>>
\begin{equation}
\label{eq:momo}
M_{0}\to M_{0}\left[\frac{\sqrt{n(T)}\,\bar{\rho}^2(T)}
{\sqrt{n(0)}\,\bar{\rho}^2(0)}\right]\equiv M_{0}(T)
\end{equation}
%EQUATION<<<
where we will use $M_{0}\approx350$ MeV as done for zero $T$. 

%-------------------------------------------------
\section{Numerical results and Discussions}
\begin{figure}[t]
\begin{tabular}{cc}
\includegraphics[width=7.5cm]{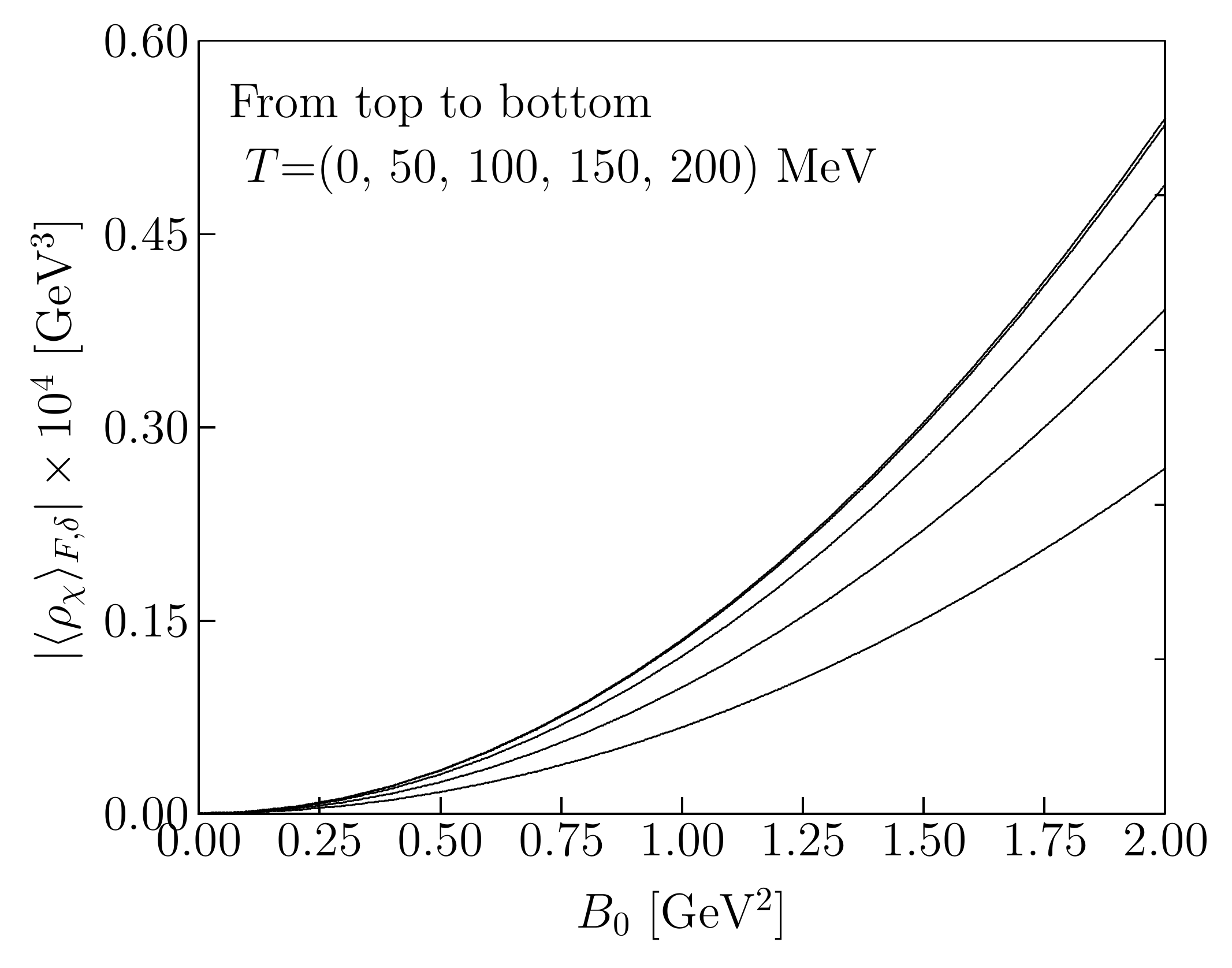}
\includegraphics[width=7.5cm]{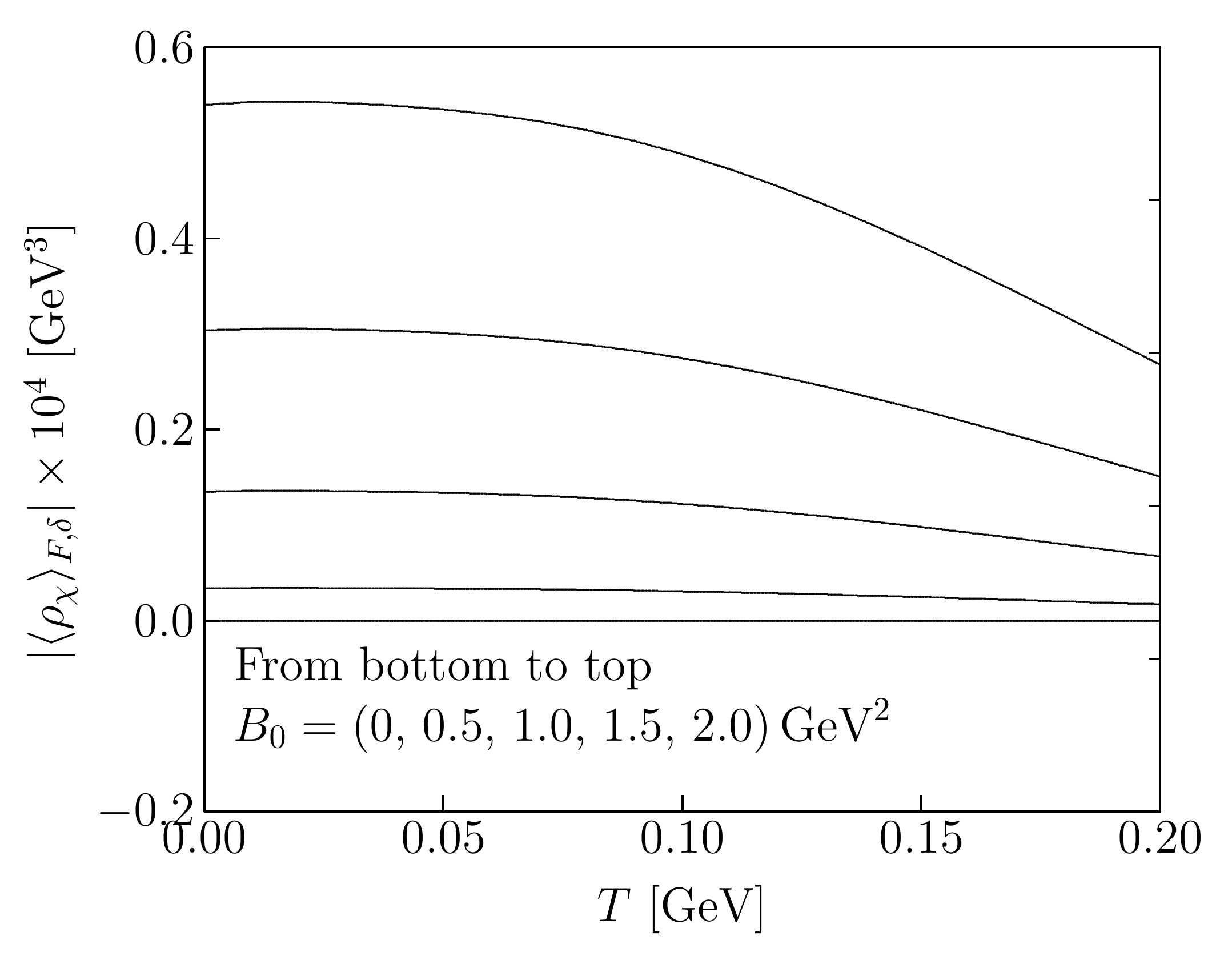}
\end{tabular}
\caption{Chiral local density, $|\langle \rho_{\chi}\rangle_{F,\delta} |\times10^{4}$, as functions of $B_{0}$ (left) and $T$ (right) for $\epsilon=10^{-3}$. The five curves in each panel correspond to those for $T=(0,50,100,150,200)$ MeV (left, from top to bottom) and $B_{0}=(0,0.5,1.0,1.5,2.0)\,\mathrm{GeV}^{2}$ (right, from bottom to top).}       
\label{FIG2}
\end{figure}

%-------------------------------------------------
First, we present the numerical results for the local chiral density $|\langle\rho_{\chi}\rangle_{F,\delta}|$ in Eq.~(\ref{eq:CD2}) as a function of $B_{0}$ in the left panel of Fig.~\ref{FIG2}. There, we depict them for different temperatures, $T=(0,\,50,\,100,\,150,\,200)$ MeV, separately. As a trial, we choose $\epsilon=10^{-3}$, which gives $\delta\approx0.0136$ in Eq.~(\ref{eq:44}). In other words, the $P$- and $CP$-violation effects are about $1\%$ in comparison to non-violating quantities. Moreover, since the $\langle \rho_{\chi}\rangle$ is almost linearly proportional to $\epsilon^{2}$, one can easily estimate it for different $\epsilon$ (or $\delta$) values. Obviously as shown in the figure, the $|\langle \rho_{\chi}\rangle|$ grows rapidly with respect to $B_{0}$, manifesting the CME. As $T$ increases, the strength of the curve decreases. At the same time, the slope of the curve, $\partial|\langle\rho_{\chi}\rangle_{F,\delta}|/\partial B_{0}$, gets smaller. This tendency means that the $|\langle\rho_{\chi}\rangle_{F,\delta}|$ becomes insensitive to the external magnetic field as $T$ gets higher, and the same for the CME. The reason for this tendency can be understood by the decreasing instanton effect.

In the right panel of Fig.~\ref{FIG2}, we show the $|\langle\rho_{\chi}\rangle_{F,\delta}|$ as a function of $T$ for different strengths of the magnetic fields, $B_{0}=(0,\,0.5,\,1.0,\,1.5,\,2.0)\,\mathrm{GeV}^{2}$. As shown there, the strength of a curve depends on that of the magnetic field. As $T$ increases, the $|\langle\rho_{\chi}\rangle_{F,\delta}|$ becomes small, indicating that the instanton effect is reduced. In other words, decreasing tunneling effect corresponding to $Q_{\mathrm{t}}\to0$. Thus, unless there is another mechanism to make the $Q_{\mathrm{t}}$ nontrivial, such as the sphaleron, the $|\langle\rho_{\chi}\rangle_{F,\delta}|$ as well as the CME decrease monotonically with respect to $T$ for a finite $B_{0}$ value. Interestingly, the $|\langle\rho_{\chi}\rangle_{F,\delta}|$ decreases much faster for the larger $B_{0}$ value, since the $T$-dependence of the quantity becomes more obvious and strengthened.

%FIGURE>>>
%FIGURE<<<

\begin{figure}[t]
\begin{tabular}{cc}
\includegraphics[width=7.5cm]{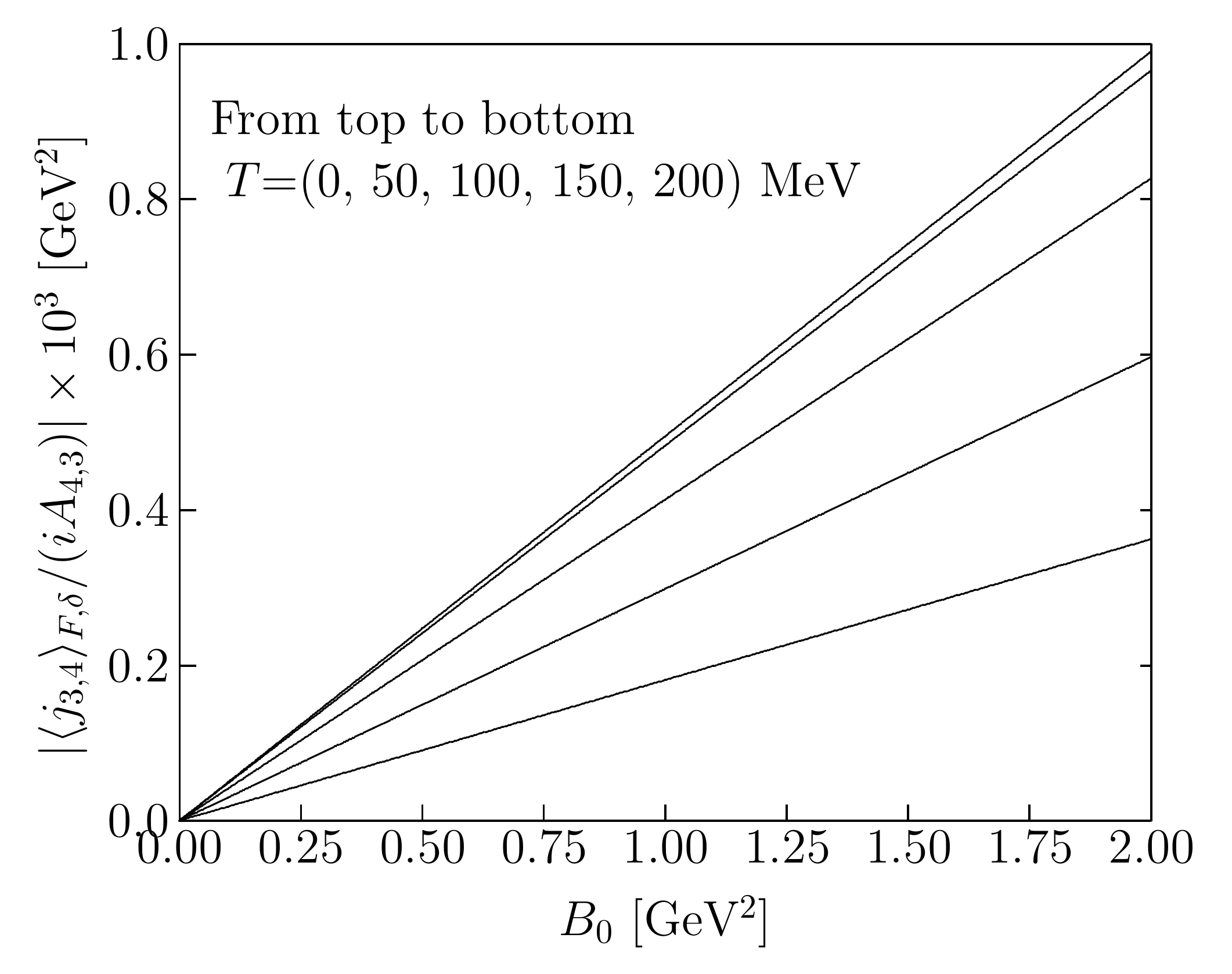}
\includegraphics[width=7.5cm]{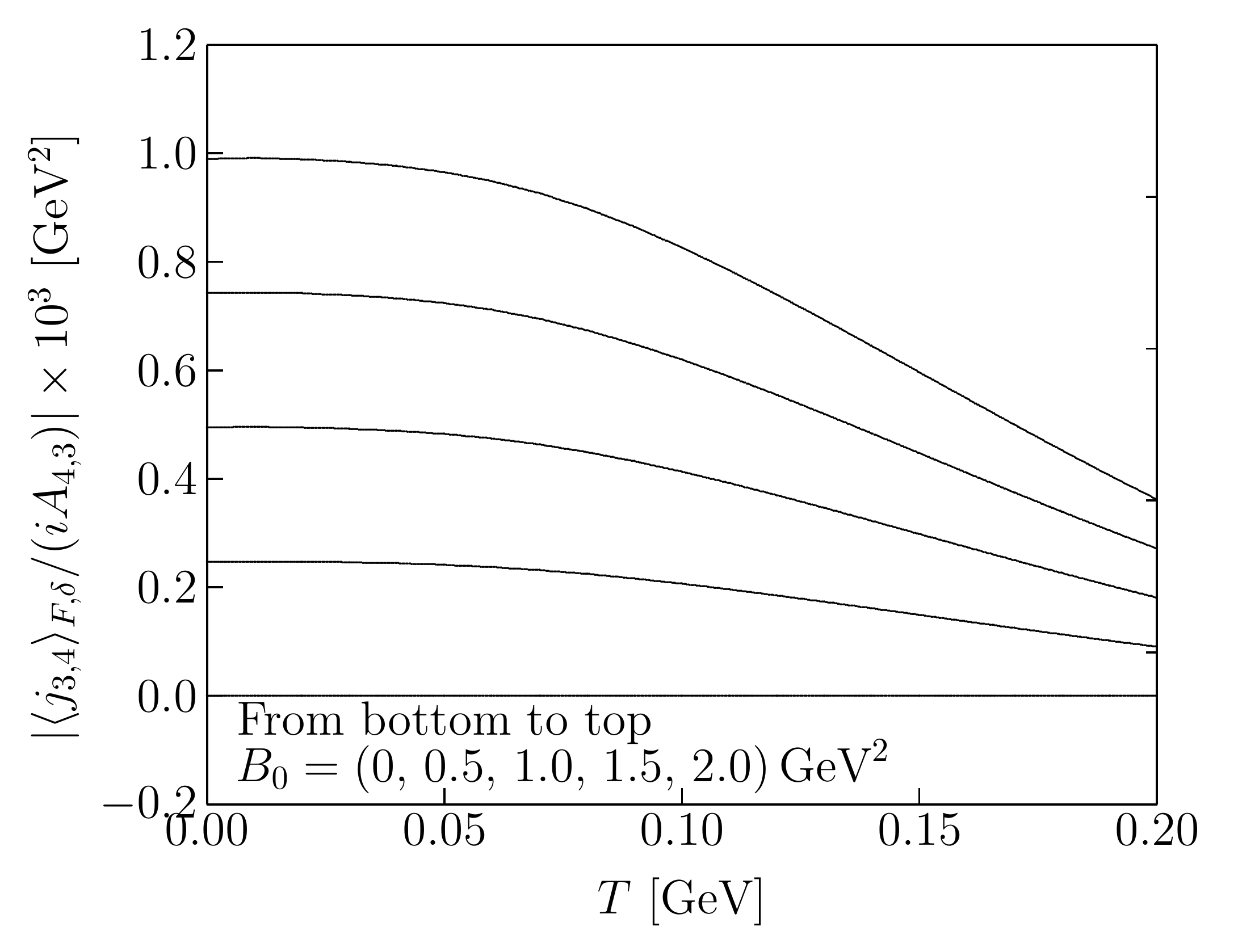}
\end{tabular}
\caption{Induced EM current indicating the CME, $|\langle j_{3,4}/(iA_{4,3})\rangle_{F,\delta}|\times10^{3}$, as functions of $B_{0}$ (left) and $T$ (right) for $\epsilon=10^{-3}$. The five curves in each panel correspond to those for $T=(0,50,100,150,200)$ MeV (left, from top to bottom) and $B_{0}=(0,0.5,1.0,1.5,2.0)\,\mathrm{GeV}^{2}$ (right, from bottom to top).}       
\label{FIG3}
\end{figure}

In the left panel of Fig.~\ref{FIG3}, we draw $|\langle j_{3,4}/(iA_{4,3})\rangle_{F,\delta}|$ as functions of $B_{0}$ for $T=(0,\,50,\,100,\,150,\,200)$ MeV, separately. The curves are increasing linearly with respect to $B_{0}$ as expected, indicating the CME. Being similar to the local chiral density, the $|\langle j_{3,4}/(iA_{4,3})\rangle_{F,\delta}|$ becomes insensitive to the magnetic field, as $T$ increases. Hence, we can conclude that the signal of the CME is weakened with respect to $T$ at a certain value of $B_{0}$. 

%FIGURE>>>
%FIGURE<<<

%-------------------------------------------------
\section{Summary and conclusion}
%-------------------------------------------------
In the present report, we have investigated the chiral magnetic effect (CME). We have observed the followings:
%ITEMIZE>>>
\begin{itemize}
\item The transverse EM current induced by the external magnetic field is far smaller than that of the longitudinal one, $j_{\parallel}\gg  j_{\perp}$, and their ratio is proportional to $\delta$, $|j_{\perp}/j_{\parallel}|\propto\delta\sim Q_{\mathrm{t}}$~\cite{Buividovich:2009wi}.
\item We obtain a relation such that the strength of the chiral charge density is the same with that of the induced EM current in the $\hat{x}_{3}$- and $\hat{x}_{4}$-directions: $\langle n_{\chi} \rangle=|\langle j_{3,4} \rangle|$ as long as the Lorentz invariance remains unbroken~\cite{Fukushima:2008xe,DHoker:1985yb}. 

\item The correct relation between the fourth component of the EM field and  the the chiral chemical potential is found: $i\delta A_{4}=\mu_{\chi}$.
\item The CME, represented by the local chiral density and induced EM current, becomes insensitive to the external magnetic field as $T$ increases. This can be understood by that the instanton, i.e.  tunneling effect gets diminished with respect to $T$: $Q_{\mathrm{t}}\to0$~\cite{Buividovich:2009wi}.
\item The local chiral density and induced EM current as functions of $T$ decrease faster for the larger external magnetic field, since the decreasing instanton effect is strengthened. 
\end{itemize} 
%ITEMIZE<<<
For more details on the present report, one can refer Ref.~\cite{Nam:2009jb}.
%-------------------------------------------------
\section*{Acknowledgment}
%-------------------------------------------------
This report is prepared for the proceeding for the international workshop {\it Hadron and Nuclear Physics} (HNP2009), 16$\sim$19 Nov 2009, Osaka, Japan. The author appreciate the hospitality during his stay at RCNP, Osaka University, Japan, and the financial support from A.~Hosaka of RCNP. He also thanks K.~Fukushima, M.~M.~Musakhanov, and C.~W.~Kao for fruitful discussions. This work was supported by the NSC96-2112-M033-003-MY3 from the National Science Council (NSC) of Taiwan. 
%--------------------------------------------------

%(^o^)
\end{document}